\documentclass{article}
\usepackage{amsmath}

\input{tcilatex}
\begin{document}

\title{Gaussian dynamics equation in normal product form}
\author{Rui He$\thanks{%
To whom correspondence should be addressed. Email: heruim@wxc.edu.cn}$ \\
%EndAName
School of Electrical and Optoelectronic Engineering, \\
West Anhui University, Lu'an, Anhui, 237012, China\\
}
\maketitle

\begin{abstract}
In this paper, we discuss the normal product form of the density operator of
multimode Gaussian states, and obtain the correlation equation between the
kernel matrix $\mathbf{R}$ of the Gaussian density operator in the normal
product form and its kernel matrix $\mathbf{G}$ in the standard quadratic
form. Further, we explore the time evolution mechanism of $\mathbf{R}$ and
obtain the Gaussian dynamical equation under the normal product $\overset{%
\cdot }{\mathbf{R}}=i(\mathbf{RJH-HJR})$. Our work is devoted to searching
for another mechanism for Gaussian dynamics. By exploring the description of
the normal ordered density matrix under the coherent state representation,
we find that our mechanism is feasible and easy to operate.

\textbf{PACS number(s):} 03.65.-w, 03.65.Ud, 03.67.-a, 42.50.Ex
\end{abstract}

\section{Introduction}

Quantum information science with continuous variable systems is developing
rapidly, presenting many exciting prospects in both its experimental
realization and theoretical research. Concepts and protocols, such as
entanglement and teleportation, initially intended only for discrete quantum
systems, have been extended to continuous variable systems, allowing more
efficient implementation and measurements. In this context, Gaussian states,
as continuous variable quantum states, play an important role in both the
experimental and theoretical fields. Gaussian states are defined as quantum
states that have Gaussian Wigner functions, while Gaussian dynamics studies
the time evolution mechanism of Gaussian state under Gaussian unitary
transformation. Two points should be paid special attention to here, one is
that the Gaussian state itself must be of Gaussian type, and the other is
that the Hamiltonian of the dynamical system in which the Gaussian state
evolves is of standard quadratic form.

There are many works on the dynamics mechanism of Gaussian state evolution
in quadratic systems \cite{r1}-\cite{r5}. However, many studies focused on
the evolution mechanism of the covariance matrix of the Gaussian state,
which almost became the paradigm of Gaussian dynamics, and most of the
research was done in this way. Here, let us make a brief introduction to
this mechanism. For a standard quadratic system, its Hamiltonian can be
written as follows%
\begin{equation}
\widehat{H}=\frac{1}{2}\widehat{A}^{T}\mathbf{H}\widehat{A},  \label{1}
\end{equation}%
where\bigskip\ $\emph{T}$ represents the transpose of the matrix and $%
\mathbf{H}$ is a positive definite, Hermitian and symmetric $2n\times 2n$
matrix, while $\widehat{A}=(\widehat{a_{1}},...,\widehat{a_{n}},\widehat{%
a_{1}}^{\dag },...,\widehat{a_{n}}^{\dag })^{T}$, in which $\widehat{a_{i}}$
and $\widehat{a_{i}}^{\dag }$ represents the creation and annihilation
operators for $n$-mode Gaussian bosonic systems, satisfying the usual
bosonic commutation relations $[\widehat{a_{i}},\widehat{a_{j}}]=[\widehat{%
a_{i}}^{\dag },\widehat{a_{j}}^{\dag }]=0$ and $[\widehat{a_{i}},\widehat{%
a_{j}}^{\dag }]=\delta _{ij}$. Then, for a Gaussian state, its
time-evolution covariance matrix $\mathbf{\sigma }(t)$ is according to the
following rules \cite{r6}%
\begin{equation}
\overset{\cdot }{\mathbf{\sigma }}(t)=\frac{d\sigma (t)}{dt}=(\mathbf{JH})%
\mathbf{\sigma +\sigma }(\mathbf{JH})^{T},  \label{2}
\end{equation}%
\bigskip where $\mathbf{J=}\left(
\begin{array}{cc}
\mathbf{0} & \mathbf{I}_{n} \\
-\mathbf{I}_{n} & \mathbf{0}%
\end{array}%
\right) $, $\mathbf{I}_{n}$ is $n\times n$ identity matrix. Thus, by solving
Eq. (2), the time evolution of the Gaussian state can be mapped as%
\begin{equation}
\mathbf{\sigma }(t)\rightarrow \mathbf{S}(t)\mathbf{\sigma }(0)\mathbf{S}%
^{T}(t).  \label{3}
\end{equation}%
Note that $\mathbf{S}(t)\equiv \exp (\mathbf{JH}t)$, which is a symplectic
matrix and satifies with%
\begin{equation}
\mathbf{S}^{T}\mathbf{JS=SJS}^{T}=\mathbf{J.}  \label{4}
\end{equation}

However, can we directly give the law of the time evolution of the Gaussian
state $\rho _{G}(t)$ itself? This is the main topic to be studied in the
present paper. In short, we give the law of the time evolution of the kernel
$\mathbf{R}$ of the Gaussian density matrix in the normal product form
through effective theoretical derivation, which is an important development
of the Gaussian dynamics mechanism. Compared with the previous work, our
work is dedicated to directly giving the time evolution of the Gaussian
density matrix, breaking the previous theoretical paradigm with the
covariance matrix as a bridge. Moreover, due to the operational simplicity
of the normal ordered operator in the coherent state representation, we can
in principle solve analytically many problems related to the evolution of
density matrices, such as the evolution of von Neumann entropy.

Our work is arranged as follows: In Sec. $2$, we first give a brief review
of the Gaussian state and its covariance matrix. Then, we use the covariance
matrix of the Gaussian state $\rho _{G}(t)$ as a bridge to obtain the
algebraic relationship between the kernel $\mathbf{G}$ of the Gaussian state
density matrix and the kernel $\mathbf{R}$ of the normal form of the density
matrix, so that once we get $\mathbf{R}$, we can give $\mathbf{G}$, vice
versa. In Sec. $3$, we introduce the coherent state representation
description of the Gaussian state, which is the basis for our follow-up
work. In Sec. $4$, we will show the time evolution law of the kernel matrix $%
\mathbf{R}$ of the normal product of $\rho _{G}(t)$

\begin{equation}
\overset{\cdot }{\mathbf{R}}=i(\mathbf{RJH-HJR}).  \label{5}
\end{equation}

\section{Gaussian state and its covariance matrix}

The density of a Gaussian state can generally be written as \cite{r7}%
\begin{equation}
\rho _{G}=\frac{e^{-\widehat{G}}}{Tr(e^{-\widehat{G}})}.  \label{6}
\end{equation}%
Note that $\widehat{G}=\frac{1}{2}\widehat{A}^{T}\mathbf{G}\widehat{A}$. By
Williamson's theorem \cite{r8}, for a positive definite, Hermitian and
symmetric $2n\times 2n$ matrix $\mathbf{G}$, it can be decomposed into the
following form%
\begin{equation}
\mathbf{G}=\mathbf{S}^{T}\widetilde{\mathbf{K}}\mathbf{S},  \label{7}
\end{equation}%
\bigskip where, $\mathbf{S}$ denotes a symplectic matrix, $\widetilde{%
\mathbf{K}}=\left(
\begin{array}{cc}
\mathbf{K} & \mathbf{0} \\
\mathbf{0} & \mathbf{K}%
\end{array}%
\right) $ and $\mathbf{K}=diag(\omega _{1},\ldots ,\omega _{n})$. According
to \cite{r7}, for the Gaussian state given by Eq. (6), its covariance matrix
can be written as%
\begin{equation}
\mathbf{\sigma }=\mathbf{S}^{-1}\widetilde{\mathbf{\nu }}\mathbf{S}^{-T},
\label{8}
\end{equation}%
in which, $\widetilde{\mathbf{\nu }}=\left(
\begin{array}{cc}
\mathbf{\nu } & \mathbf{0} \\
\mathbf{0} & \mathbf{\nu }%
\end{array}%
\right) $, $\mathbf{\nu }=diag(\nu _{1},\ldots ,\nu _{n})$, and $\nu _{i}=%
\frac{1+e^{-\omega _{i}}}{1-e^{-\omega _{i}}}$. Then%
\begin{equation}
\mathbf{\sigma }=\frac{\mathbf{I}+e^{-\mathbf{\Omega G}}}{\mathbf{I}-e^{-%
\mathbf{\Omega G}}}\mathbf{\Omega =\coth (}\frac{\mathbf{\Omega G}}{2})%
\mathbf{\Omega ,}  \label{9}
\end{equation}%
where, $\mathbf{\Omega }=\left(
\begin{array}{cc}
\mathbf{I}_{n} & \mathbf{0} \\
\mathbf{0} & \mathbf{-I}_{n}%
\end{array}%
\right) $.

We also know that the characteristic function of any Gaussian state can be
written as \cite{r9}%
\begin{equation}
C(\mathbf{Z})=e^{-\frac{1}{2}\mathbf{Z}^{\dag }\mathbf{CZ}}.  \label{10}
\end{equation}%
Note that $\mathbf{Z=(}z_{1},\ldots ,z_{n},z_{1}^{\ast },\ldots ,z_{n}^{\ast
})^{T}$. By using

\begin{equation}
e^{\mathbf{Z}^{\dag }\mathbf{\Omega }\widehat{A}}=\colon e^{\mathbf{Z}^{\dag
}\mathbf{\Omega }\widehat{A}-\frac{1}{4}\mathbf{Z}^{\dag }\mathbf{Z}}\colon ,
\label{11}
\end{equation}%
where, $\colon \cdots \colon $ represents normal ordering. Then,

\begin{eqnarray}
\rho _{G} &=&\dint (d\mathbf{Z})e^{\mathbf{Z}^{\dag }\mathbf{\Omega }%
\widehat{A}}C(\mathbf{Z})  \label{12} \\
&=&\dint (d\mathbf{Z})\colon e^{\mathbf{Z}^{\dag }\mathbf{\Omega }\widehat{A}%
-\frac{1}{4}\mathbf{Z}^{\dag }\mathbf{Z}}\colon e^{-\frac{1}{2}\mathbf{Z}%
^{\dag }\mathbf{CZ}}  \notag \\
&=&\dint (d\mathbf{Z})\colon e^{-\frac{1}{2}\mathbf{Z}^{\dag }\mathbf{(C+}%
\frac{1}{2}\mathbf{I)Z}}e^{\mathbf{Z}^{\dag }\mathbf{\Omega }\widehat{A}%
}\colon .  \notag
\end{eqnarray}%
By using the technique of integration within ordered product (IWOP) \cite%
{r10} and the integeral fomula%
\begin{equation}
\dint (d\mathbf{Z})e^{-\frac{1}{2}\mathbf{Z}^{\dag }\mathbf{VZ}}e^{\mathbf{Z}%
^{\dag }\mathbf{X}}=\frac{1}{\sqrt{\det \mathbf{V}}}e^{-\frac{1}{2}\mathbf{X}%
^{T}\mathbf{EV}^{-1}\mathbf{X}},  \label{13}
\end{equation}%
where, $\mathbf{E=}\left(
\begin{array}{cc}
\mathbf{0} & \mathbf{I}_{n} \\
\mathbf{I}_{n} & \mathbf{0}%
\end{array}%
\right) ,$ let us continue our derivation
\begin{eqnarray}
\rho _{G} &=&\frac{1}{\sqrt{\det \mathbf{(C+}\frac{1}{2}\mathbf{I)}}}\colon
\exp [-\frac{1}{2}\mathbf{(\Omega }\widehat{A}\mathbf{)}^{T}\mathbf{E(C+}%
\frac{1}{2}\mathbf{I)}^{-1}(\mathbf{\Omega }\widehat{A})]\colon  \label{14}
\\
&=&\frac{1}{\sqrt{\det \mathbf{(C+}\frac{1}{2}\mathbf{I)}}}\colon \exp [-%
\frac{1}{2}\widehat{A}^{T}\mathbf{\Omega E(C+}\frac{1}{2}\mathbf{I)}^{-1}%
\mathbf{\Omega }\widehat{A}]\colon .  \notag
\end{eqnarray}%
Here, we can set $\mathbf{R\equiv \Omega E(C+}\frac{1}{2}\mathbf{I)}^{-1}%
\mathbf{\Omega }$\textbf{, }then\textbf{\ }%
\begin{equation}
\rho _{G}=\sqrt{\det \mathbf{R}}\colon \exp (-\frac{1}{2}\widehat{A}^{T}%
\mathbf{R}\widehat{A})\colon .  \label{15}
\end{equation}

Since the Wigner function of the Gaussian state $\rho _{G}$ can be written as%
\begin{equation}
W(\mathbf{Z})=\frac{1}{\sqrt{\det \mathbf{\sigma }}}\exp (-\mathbf{Z}^{\dag }%
\mathbf{\sigma }^{-1}\mathbf{Z).}  \label{16}
\end{equation}%
Note that $\mathbf{\sigma }$ here is the covariance matrix in Eq. (2).
According to the Fourier transform relationship between $C(\mathbf{Z})$ and $%
W(\mathbf{Z})$, we can get

\begin{equation}
\frac{\mathbf{\sigma }^{-1}}{2}=\mathbf{\Omega C}^{-1}\mathbf{\Omega }
\label{17}
\end{equation}%
or%
\begin{equation}
\mathbf{C}=\frac{1}{2}\mathbf{\Omega \sigma \Omega .}  \label{18}
\end{equation}%
Substituting Eq. (9) into Eq. (18), we have%
\begin{equation}
\mathbf{C}=\frac{\mathbf{\Omega }}{2}\frac{\mathbf{I}+e^{-\mathbf{\Omega G}}%
}{\mathbf{I}-e^{-\mathbf{\Omega G}}}.  \label{19}
\end{equation}%
Then, taking Eq. (19) into Eq. (14), we can get%
\begin{eqnarray}
\mathbf{R}\mathbf{=\Omega E(} &&\frac{\mathbf{I}}{2}\mathbf{+}\frac{\mathbf{%
\Omega }}{2}\frac{\mathbf{I}+e^{-\mathbf{\Omega G}}}{\mathbf{I}-e^{-\mathbf{%
\Omega G}}}\mathbf{)}^{-1}\mathbf{\Omega }  \label{20} \\
&=&-2\mathbf{E\Omega (I+\Omega }\frac{\mathbf{I}+e^{-\mathbf{\Omega G}}}{%
\mathbf{I}-e^{-\mathbf{\Omega G}}}\mathbf{)}^{-1}\mathbf{\Omega }  \notag \\
&=&-2\mathbf{E(I+}\frac{\mathbf{I}+e^{-\mathbf{\Omega G}}}{\mathbf{I}-e^{-%
\mathbf{\Omega G}}}\mathbf{\Omega )}^{-1}  \notag \\
&=&-2\mathbf{(E+}\frac{\mathbf{I}+e^{-\mathbf{\Omega G}}}{\mathbf{I}-e^{-%
\mathbf{\Omega G}}}\mathbf{\Omega E)}^{-1}  \notag \\
&=&-2\mathbf{(E+}\frac{\mathbf{I}+e^{-\mathbf{\Omega G}}}{\mathbf{I}-e^{-%
\mathbf{\Omega G}}}\mathbf{J)}^{-1}  \notag \\
&=&-2\mathbf{(E+JJ}^{-1}\frac{\mathbf{I}+e^{-\mathbf{\Omega G}}}{\mathbf{I}%
-e^{-\mathbf{\Omega G}}}\mathbf{J)}^{-1}  \notag \\
&=&-2\mathbf{(E+J}\frac{\mathbf{I}+e^{-\mathbf{J}^{-1}\mathbf{\Omega GJ}}}{%
\mathbf{I}-e^{-\mathbf{J}^{-1}\mathbf{\Omega GJ}}}\mathbf{)}^{-1}  \notag \\
&=&-2\mathbf{(E+J}\frac{\mathbf{I}+e^{-\mathbf{EGJ}}}{\mathbf{I}-e^{-\mathbf{%
EGJ}}}\mathbf{)}^{-1}.  \notag
\end{eqnarray}%
In this way, we obtain the relationship of the kernel matrix $\mathbf{R}$ of
the normal product of $\rho _{G}(t)$ and $\mathbf{G}$, which is exactly the
same results as in \cite{r11}. In Gaussian dynamics, as long as we know the
time evolution of $\mathbf{R}$, we can infer the evolution of $\mathbf{G}$
from Eq. (20). That is to say, we can directly calculate the time evolution
of the density matrix of the Gaussian state by using this method. Moreover,
according to the above calculation, we can also deduce the relationship
between $\mathbf{R}$ and $\mathbf{\sigma }$

\begin{equation}
\mathbf{R}=-2\mathbf{E}(\mathbf{\sigma +I})^{-1}.  \label{21}
\end{equation}

\section{Coherent state representation of Gaussian state}

Now we introduce $n$-mode coherent states $|\mathbf{Z}\rangle \equiv
|z_{1},...,z_{n}\rangle $ and suppose that $\rho (\mathbf{Z})=\langle
\mathbf{Z}|\rho _{G}|\mathbf{Z}\rangle $. In normal product form, bosonic
creation and annihilation operators could be replaced by the complex
parameter of the coherent state, thus, we have
\begin{eqnarray}
\rho (\mathbf{Z}) &=&\sqrt{\det \mathbf{R}}\langle \mathbf{Z}|\colon \exp (-%
\frac{1}{2}\widehat{A}^{T}\mathbf{R}\widehat{A})\colon |\mathbf{Z}\rangle
\label{22} \\
&=&\sqrt{\det \mathbf{R}}e^{-\frac{1}{2}\mathbf{Z}^{T}\mathbf{RZ}}.  \notag
\end{eqnarray}%
For a single-mode coherent state $|z\rangle $, we have%
\begin{equation}
|z\rangle \langle z|\widehat{a}=(z+\frac{\partial }{\partial z^{\ast }}%
)|z\rangle \langle z|,  \label{23}
\end{equation}%
\begin{equation}
\widehat{a}^{\dag }|z\rangle \langle z|=(z^{\ast }+\frac{\partial }{\partial
z})|z\rangle \langle z|.  \label{24}
\end{equation}%
We can generalize the relationship given by the above two equations to the
multimode case and have

\begin{eqnarray}
\widehat{A}|\mathbf{Z}\rangle \langle \mathbf{Z}| &=&\left(
\begin{array}{c}
\widehat{a}_{1} \\
\vdots \\
\widehat{a}_{n} \\
\widehat{a}_{1}^{\dag } \\
\vdots \\
\widehat{a}_{n}^{\dag }%
\end{array}%
\right) |\mathbf{Z}\rangle \langle \mathbf{Z}|=\left[ \left(
\begin{array}{c}
z_{1} \\
\vdots \\
z_{n} \\
z_{1}^{\ast } \\
\vdots \\
z_{n}^{\ast }%
\end{array}%
\right) +\left(
\begin{array}{c}
0 \\
\vdots \\
0 \\
\frac{\partial }{\partial z_{1}} \\
\vdots \\
\frac{\partial }{\partial z_{n}}%
\end{array}%
\right) \right] |\mathbf{Z}\rangle \langle \mathbf{Z}|  \label{25} \\
&=&(\mathbf{Z}+\frac{\mathbf{E-J}}{2}\frac{\partial }{\partial \mathbf{Z}^{T}%
})|\mathbf{Z}\rangle \langle \mathbf{Z}|.  \notag
\end{eqnarray}%
Similarly, the following formula can be derived%
\begin{equation}
|\mathbf{Z}\rangle \langle \mathbf{Z}|\widehat{A}^{T}=(\mathbf{Z}^{T}+\frac{%
\mathbf{E+J}}{2}\frac{\partial }{\partial \mathbf{Z}})|\mathbf{Z}\rangle
\langle \mathbf{Z}|.  \label{26}
\end{equation}%
Taking into account Eqs. (25) and (26), in the coherent state
representation, we obtain

\begin{eqnarray}
\langle \mathbf{Z}|\rho _{G}\widehat{A}|\mathbf{Z}\rangle &=&\langle \mathbf{%
Z}|\rho _{G}|\mathbf{Z}\rangle (\mathbf{Z}+\overleftarrow{\frac{\partial }{%
\partial \mathbf{Z}^{T}}}\frac{\mathbf{E-J}}{2})  \label{27} \\
&=&\rho (\mathbf{Z})(\mathbf{Z}+\overleftarrow{\frac{\partial }{\partial
\mathbf{Z}^{T}}}\frac{\mathbf{E-J}}{2})  \notag
\end{eqnarray}%
and%
\begin{eqnarray}
\langle \mathbf{Z}|\widehat{A}^{T}\rho _{G}|\mathbf{Z}\rangle &=&(\mathbf{Z}%
^{T}+\frac{\mathbf{E+J}}{2}\frac{\partial }{\partial \mathbf{Z}})\langle
\mathbf{Z}|\rho _{G}|\mathbf{Z}\rangle  \label{28} \\
&=&(\mathbf{Z}^{T}+\frac{\mathbf{E+J}}{2}\frac{\partial }{\partial \mathbf{Z}%
})\rho (\mathbf{Z}),  \notag
\end{eqnarray}%
where, we have set $\rho (\mathbf{Z})\equiv \langle \mathbf{Z}|\rho _{G}|%
\mathbf{Z}\rangle $, which is actually a Husimi-Q function in the phase
space representation.

\section{Gaussian dynamics equation in normal product form}

For an open dynamic system, the time evolution mechanism of the system is
determined by the following Lindblad equation \cite{r12}

\begin{equation}
\overset{\cdot }{\rho }(t)=-i[\widehat{H},\rho (t)]+\underset{i}{\dsum }[%
\widehat{c_{i}}\rho (t)\widehat{c_{i}}^{\dag }-\frac{1}{2}\widehat{c_{i}}%
^{\dag }\widehat{c_{i}}\rho (t)-\frac{1}{2}\rho (t)\widehat{c_{i}}^{\dag }%
\widehat{c_{i}}],  \label{29}
\end{equation}%
where $\widehat{H}$ is quadratic, $\widehat{c_{i}}$ and $\widehat{c_{i}}%
^{\dag }$ are the linear forms of the creation and annihilation operators.
Although the content discussed in this paper can be fully extended to the
case where the quantum system is affected by the coherent environment, that
is, considering the second term on the right side of Eq. (29), for the sake
of brevity and beauty of the text, we only analyze the time evolution
mechanism of Gaussian states in quadratic Hamiltonian systems independent of
the environment. That is to say, we only discuss the quantum Liouville
equation

\begin{equation}
\overset{\cdot }{\rho _{G}}(t)=i[\rho _{G}(t),\widehat{H}].  \label{30}
\end{equation}%
Note that here $\widehat{H}=\frac{1}{2}\widehat{A}^{T}\mathbf{H}\widehat{A}$
and $\rho _{G}=\frac{e^{-\widehat{G}}}{Tr(e^{-\widehat{G}})}=\sqrt{\det
\mathbf{R}}\colon \exp (-\frac{1}{2}\widehat{A}^{T}\mathbf{R}\widehat{A}%
)\colon $. Substituting $\widehat{H}$ and $\rho _{G}$\ into Eq. (30), we get

\begin{equation}
\frac{d[\sqrt{\det \mathbf{R}}\colon \exp (-\frac{1}{2}\widehat{A}^{T}%
\mathbf{R}\widehat{A})\colon ]}{dt}=-\frac{i}{2}\sqrt{\det \mathbf{R}}[%
\widehat{A}^{T}\mathbf{H}\widehat{A},\colon \exp (-\frac{1}{2}\widehat{A}^{T}%
\mathbf{R}\widehat{A})\colon ].  \label{31}
\end{equation}%
By using the commutation formula $[AB,C]=A[B,C]+[A,C]B$, we obtain%
\begin{eqnarray}
&&\frac{d[\sqrt{\det \mathbf{R}}\colon \exp (-\frac{1}{2}\widehat{A}^{T}%
\mathbf{R}\widehat{A})\colon ]}{dt}  \label{32} \\
&=&-\frac{i}{2}\sqrt{\det \mathbf{R}}\widehat{A}^{T}\mathbf{H}[\widehat{A}%
,\colon \exp (-\frac{1}{2}\widehat{A}^{T}\mathbf{R}\widehat{A})\colon ]-%
\frac{i}{2}\sqrt{\det \mathbf{R}}\widehat{A}^{T}[\mathbf{H},\colon \exp (-%
\frac{1}{2}\widehat{A}^{T}\mathbf{R}\widehat{A})\colon ]\widehat{A}  \notag
\\
&&-\frac{i}{2}\sqrt{\det \mathbf{R}}[\widehat{A}^{T},\colon \exp (-\frac{1}{2%
}\widehat{A}^{T}\mathbf{R}\widehat{A})\colon ]\mathbf{H}\widehat{A}.  \notag
\end{eqnarray}%
Considering the following normal product properties \cite{r13}

\begin{equation}
\colon \frac{\partial }{\partial \widehat{a}}f(\widehat{a},\widehat{a}^{\dag
})\colon =[\colon f(\widehat{a},\widehat{a}^{\dag })\colon ,\widehat{a}%
^{\dag }],  \label{33}
\end{equation}

\begin{equation}
\colon \frac{\partial }{\partial \widehat{a}^{\dag }}f(\widehat{a},\widehat{a%
}^{\dag })\colon =[\widehat{a},\colon f(\widehat{a},\widehat{a}^{\dag
})\colon ],  \label{34}
\end{equation}%
and the derivation rule of quadratic matrix

\begin{equation}
\frac{d(X^{T}AX)}{dX}=2X^{T}A,  \label{35}
\end{equation}

\begin{equation}
\frac{d(X^{T}AX)}{dX^{T}}=2AX,  \label{36}
\end{equation}%
under the condition $A=A^{T}$ ($A$ is a symmetric matrix), we can simplify
Eq. (32) into the following form

\begin{eqnarray}
&&\frac{d[\sqrt{\det \mathbf{R}}\colon \exp (-\frac{1}{2}\widehat{A}^{T}%
\mathbf{R}\widehat{A})\colon ]}{dt}  \label{37} \\
&=&-\frac{i}{2}\sqrt{\det \mathbf{R}}\widehat{A}^{T}\mathbf{H}\colon \mathbf{%
J}\frac{\partial }{\partial \widehat{A}^{T}}\exp (-\frac{1}{2}\widehat{A}^{T}%
\mathbf{R}\widehat{A})\colon +\frac{i}{2}\sqrt{\det \mathbf{R}}\colon \frac{%
\partial }{\partial \widehat{A}}\exp (-\frac{1}{2}\widehat{A}^{T}\mathbf{R}%
\widehat{A})\mathbf{J}\colon \mathbf{H}\widehat{A}  \notag \\
&&-\frac{i}{2}\sqrt{\det \mathbf{R}}\widehat{A}^{T}[\mathbf{H},\colon \exp (-%
\frac{1}{2}\widehat{A}^{T}\mathbf{R}\widehat{A})\colon ]\widehat{A}  \notag
\\
&=&\frac{i}{2}\sqrt{\det \mathbf{R}}\widehat{A}^{T}\mathbf{H}\colon \mathbf{%
JR}\widehat{A}\exp (-\frac{1}{2}\widehat{A}^{T}\mathbf{R}\widehat{A})\colon -%
\frac{i}{2}\sqrt{\det \mathbf{R}}\colon \widehat{A}^{T}\mathbf{R}\exp (-%
\frac{1}{2}\widehat{A}^{T}\mathbf{R}\widehat{A})\mathbf{J}\colon \mathbf{H}%
\widehat{A}  \notag \\
&&-\frac{i}{2}\sqrt{\det \mathbf{R}}\widehat{A}^{T}[\mathbf{H},\colon \exp (-%
\frac{1}{2}\widehat{A}^{T}\mathbf{R}\widehat{A})\colon ]\widehat{A}.  \notag
\end{eqnarray}%
In the dynamics of phase space, the time evolution formula of Husimi-Q
function $\rho (\mathbf{Z})$ can be derived as follow

\begin{eqnarray}
\frac{d\rho (\mathbf{Z})}{dt} &=&Tr(\overset{\cdot }{\rho }|\mathbf{Z}%
\rangle \langle \mathbf{Z}|)  \label{38} \\
&=&-iTr(\rho \widehat{H}|\mathbf{Z}\rangle \langle \mathbf{Z}|-\widehat{H}%
\rho |\mathbf{Z}\rangle \langle \mathbf{Z}|)  \notag \\
&=&-i\langle \mathbf{Z}|\rho \widehat{H}|\mathbf{Z}\rangle +i\langle \mathbf{%
Z}|\widehat{H}\rho |\mathbf{Z}\rangle ,  \notag
\end{eqnarray}%
In fact, we just need to average the coherent states on both sides of the
Liouville equation. By calculating the average value of the coherent states
on both sides of Eq. (37), we have

\begin{eqnarray}
&&\frac{d[\sqrt{\det \mathbf{R}}\langle \mathbf{Z}|\colon \exp (-\frac{1}{2}%
\widehat{A}^{T}\mathbf{R}\widehat{A})\colon |\mathbf{Z}\rangle ]}{dt}
\label{39} \\
&=&\frac{i}{2}\sqrt{\det \mathbf{R}}\langle \mathbf{Z}|\widehat{A}^{T}%
\mathbf{HJR}\colon \exp (-\frac{1}{2}\widehat{A}^{T}\mathbf{R}\widehat{A})%
\widehat{A}\colon |\mathbf{Z}\rangle  \notag \\
&&-\frac{i}{2}\sqrt{\det \mathbf{R}}\langle \mathbf{Z}|\colon \widehat{A}%
^{T}\exp (-\frac{1}{2}\widehat{A}^{T}\mathbf{R}\widehat{A})\colon \mathbf{RJH%
}\widehat{A}|\mathbf{Z}\rangle  \notag \\
&&-\frac{i}{2}\sqrt{\det \mathbf{R}}\langle \mathbf{Z}|\widehat{A}^{T}[%
\mathbf{H},\colon \exp (-\frac{1}{2}\widehat{A}^{T}\mathbf{R}\widehat{A}%
)\colon ]\widehat{A}|\mathbf{Z}\rangle .  \notag
\end{eqnarray}%
We first calculate the third part of the right-hand side of Eq. (38) and have

\begin{eqnarray}
&&-\frac{i}{2}\sqrt{\det \mathbf{R}}\langle \mathbf{Z}|\widehat{A}^{T}[%
\mathbf{H},\colon \exp (-\frac{1}{2}\widehat{A}^{T}\mathbf{R}\widehat{A}%
)\colon ]\widehat{A}|\mathbf{Z}\rangle  \label{40} \\
&=&-\frac{i}{2}\sqrt{\det \mathbf{R}}(\mathbf{Z}^{T}+\frac{\mathbf{E+J}}{2}%
\frac{\partial }{\partial \mathbf{Z}})\langle \mathbf{Z}|[\mathbf{H},\colon
\exp (-\frac{1}{2}\widehat{A}^{T}\mathbf{R}\widehat{A})\colon ]|\mathbf{Z}%
\rangle (\mathbf{Z}+\overleftarrow{\frac{\partial }{\partial \mathbf{Z}^{T}}}%
\frac{\mathbf{E-J}}{2})  \notag \\
&=&-\frac{i}{2}\sqrt{\det \mathbf{R}}(\mathbf{Z}^{T}+\frac{\mathbf{E+J}}{2}%
\frac{\partial }{\partial \mathbf{Z}})(\langle \mathbf{Z}|\mathbf{H}\colon
\exp (-\frac{1}{2}\widehat{A}^{T}\mathbf{R}\widehat{A})\colon |\mathbf{Z}%
\rangle  \notag \\
&&-\langle \mathbf{Z}|\colon \exp (-\frac{1}{2}\widehat{A}^{T}\mathbf{R}%
\widehat{A})\colon \mathbf{H}|\mathbf{Z}\rangle )(\mathbf{Z}+\overleftarrow{%
\frac{\partial }{\partial \mathbf{Z}^{T}}}\frac{\mathbf{E-J}}{2})  \notag \\
&=&-\frac{i}{2}\sqrt{\det \mathbf{R}}(\mathbf{Z}^{T}+\frac{\mathbf{E+J}}{2}%
\frac{\partial }{\partial \mathbf{Z}})(\mathbf{H}\langle \mathbf{Z}|\colon
\exp (-\frac{1}{2}\widehat{A}^{T}\mathbf{R}\widehat{A})\colon |\mathbf{Z}%
\rangle  \notag \\
&&-\langle \mathbf{Z}|\colon \exp (-\frac{1}{2}\widehat{A}^{T}\mathbf{R}%
\widehat{A})\colon |\mathbf{Z}\rangle \mathbf{H})(\mathbf{Z}+\overleftarrow{%
\frac{\partial }{\partial \mathbf{Z}^{T}}}\frac{\mathbf{E-J}}{2})  \notag \\
&=&-\frac{i}{2}(\mathbf{Z}^{T}+\frac{\mathbf{E+J}}{2}\frac{\partial }{%
\partial \mathbf{Z}})(\mathbf{H}\rho (\mathbf{Z})-\rho (\mathbf{Z})\mathbf{H)%
}(\mathbf{Z}+\overleftarrow{\frac{\partial }{\partial \mathbf{Z}^{T}}}\frac{%
\mathbf{E-J}}{2}).  \notag
\end{eqnarray}%
Since $\rho (\mathbf{Z})$ is a number, $\mathbf{H}\rho (\mathbf{Z})-\rho (%
\mathbf{Z})\mathbf{H=0}$. So we show $-\frac{i}{2}\sqrt{\det \mathbf{R}}%
\langle \mathbf{Z}|\widehat{A}^{T}[\mathbf{H},\colon \exp (-\frac{1}{2}%
\widehat{A}^{T}\mathbf{R}\widehat{A})\colon ]\widehat{A}|\mathbf{Z}\rangle
=0 $. We continue to calculate the first two terms on the right-hand side of
Eq. (39),

\begin{eqnarray}
&&\frac{i}{2}\sqrt{\det \mathbf{R}}\langle \mathbf{Z}|\widehat{A}^{T}\mathbf{%
HJR}\colon \exp (-\frac{1}{2}\widehat{A}^{T}\mathbf{R}\widehat{A})\widehat{A}%
\colon |\mathbf{Z}\rangle  \label{41} \\
&=&\frac{i}{2}\sqrt{\det \mathbf{R}}(\mathbf{Z}^{T}+\frac{\mathbf{E+J}}{2}%
\frac{\partial }{\partial \mathbf{Z}})[\mathbf{HJR}\langle \mathbf{Z}|\colon
\exp (-\frac{1}{2}\widehat{A}^{T}\mathbf{R}\widehat{A})\widehat{A}\colon |%
\mathbf{Z}\rangle \mathbf{]}  \notag
\end{eqnarray}%
and

\begin{eqnarray}
&&-\frac{i}{2}\sqrt{\det \mathbf{R}}\langle \mathbf{Z}|\colon \widehat{A}%
^{T}\exp (-\frac{1}{2}\widehat{A}^{T}\mathbf{R}\widehat{A})\colon \mathbf{RJH%
}\widehat{A}|\mathbf{Z}\rangle  \label{42} \\
&=&-\frac{i}{2}\sqrt{\det \mathbf{R}}\langle \mathbf{Z}|\colon \widehat{A}%
^{T}\exp (-\frac{1}{2}\widehat{A}^{T}\mathbf{R}\widehat{A})\colon \mathbf{RJH%
}|\mathbf{Z}\rangle (\mathbf{Z}+\overleftarrow{\frac{\partial }{\partial
\mathbf{Z}^{T}}}\frac{\mathbf{E-J}}{2}).  \notag
\end{eqnarray}%
Then,

\begin{eqnarray}
\frac{d\mathbf{\rho (\mathbf{Z})}}{dt} &=&\frac{d[\sqrt{\det \mathbf{R}}%
\langle \mathbf{Z}|\colon \exp (-\frac{1}{2}\widehat{A}^{T}\mathbf{R}%
\widehat{A})\colon |\mathbf{Z}\rangle ]}{dt}  \label{43} \\
&=&-\frac{1}{2}\sqrt{\det \mathbf{R}}\mathbf{Z}^{T}\overset{\cdot }{\mathbf{R%
}}\mathbf{Z}\overset{\symbol{126}}{\mathbf{\rho }}\mathbf{(\mathbf{Z})+}%
\frac{d\sqrt{\det \mathbf{R}}}{dt}\overset{\symbol{126}}{\mathbf{\rho }}%
\mathbf{(\mathbf{Z})}  \notag \\
&=&\frac{i}{2}\sqrt{\det \mathbf{R}}(\mathbf{Z}^{T}+\frac{\mathbf{E+J}}{2}%
\frac{\partial }{\partial \mathbf{Z}})[\mathbf{HJR}\langle \mathbf{Z}|\colon
\exp (-\frac{1}{2}\widehat{A}^{T}\mathbf{R}\widehat{A})\widehat{A}\colon |%
\mathbf{Z}\rangle \mathbf{]}  \notag \\
&&-\frac{i}{2}\sqrt{\det \mathbf{R}}\langle \mathbf{Z}|\colon \widehat{A}%
^{T}\exp (-\frac{1}{2}\widehat{A}^{T}\mathbf{R}\widehat{A})\colon \mathbf{RJH%
}|\mathbf{Z}\rangle (\mathbf{Z}+\overleftarrow{\frac{\partial }{\partial
\mathbf{Z}^{T}}}\frac{\mathbf{E-J}}{2})  \notag \\
&=&\frac{i}{2}\sqrt{\det \mathbf{R}}(\mathbf{Z}^{T}+\frac{\mathbf{E+J}}{2}%
\frac{\partial }{\partial \mathbf{Z}})[\mathbf{HJR}\overset{\symbol{126}}{%
\mathbf{\rho }}(\mathbf{Z})\mathbf{Z]}  \notag \\
&&-\frac{i}{2}\sqrt{\det \mathbf{R}}[\mathbf{Z}^{T}\overset{\symbol{126}}{%
\mathbf{\rho }}(\mathbf{Z})\mathbf{RJH}](\mathbf{Z}+\overleftarrow{\frac{%
\partial }{\partial \mathbf{Z}^{T}}}\frac{\mathbf{E-J}}{2})  \notag \\
&=&\frac{i}{2}\sqrt{\det \mathbf{R}}\mathbf{Z}^{T}(\mathbf{HJR-RJH)Z}\overset%
{\symbol{126}}{\mathbf{\rho }}\mathbf{(\mathbf{Z})}+\frac{i}{2}\sqrt{\det
\mathbf{R}}\frac{\mathbf{E+J}}{2}\frac{\partial }{\partial \mathbf{Z}}[%
\mathbf{HJR\overset{\symbol{126}}{\mathbf{\rho }}\mathbf{(\mathbf{Z})}Z]}
\notag \\
&&-\frac{i}{2}\sqrt{\det \mathbf{R}}[\mathbf{Z}^{T}\overset{\symbol{126}}{%
\mathbf{\rho }}\mathbf{(\mathbf{Z})RJH]}\overleftarrow{\frac{\partial }{%
\partial \mathbf{Z}^{T}}}\frac{\mathbf{E-J}}{2}).  \notag
\end{eqnarray}%
Note that here we have set $\overset{\symbol{126}}{\mathbf{\rho }}\mathbf{(%
\mathbf{Z})=\rho (\mathbf{Z})/}\sqrt{\det \mathbf{R}}$. Multipling $\mathbf{%
E+J}$ on the left-hand side of Eq. (43) and $\mathbf{E-J}$ on its right-hand
side and noting that $\left( \mathbf{E+J}\right) ^{2}=0$ and $\left( \mathbf{%
E-J}\right) ^{2}=0$, we obtain

\begin{eqnarray}
&&\left( \mathbf{E+J}\right) \mathbf{Z}^{T}\overset{\cdot }{\mathbf{R}}%
\mathbf{Z\left( \mathbf{E-J}\right) -2\left( \mathbf{E+J}\right) }\frac{1}{%
\sqrt{\det \mathbf{R}}}\mathbf{\frac{d\sqrt{\det \mathbf{R}}}{dt}\mathbf{%
\left( \mathbf{E-J}\right) }}  \label{44} \\
&\mathbf{=}&-i\left( \mathbf{E+J}\right) \mathbf{Z}^{T}\mathbf{(HJR-RJH)Z}%
\left( \mathbf{E-J}\right) .  \notag
\end{eqnarray}%
Because $\mathbf{Z}^{T}\overset{\cdot }{\mathbf{R}}\mathbf{Z}$ , $\frac{d%
\sqrt{\det \mathbf{R}}}{dt}$and $\mathbf{Z}^{T}\mathbf{(HJR-RJH)Z}$ are all
numbers, Eq. (44) can be written as

\begin{eqnarray}
&&\left( \mathbf{E+J}\right) \mathbf{\left( \mathbf{E-J}\right) Z}^{T}%
\overset{\cdot }{\mathbf{R}}\mathbf{Z-2\left( \mathbf{E+J}\right) \mathbf{%
\left( \mathbf{E-J}\right) }\frac{d\ln \sqrt{\det \mathbf{R}}}{dt}}
\label{45} \\
&\mathbf{=}&-i\left( \mathbf{E+J}\right) \left( \mathbf{E-J}\right) \mathbf{Z%
}^{T}\mathbf{(HJR-RJH)Z.}  \notag
\end{eqnarray}%
Obviously, we have

\begin{equation}
\mathbf{Z}^{T}[\overset{\cdot }{\mathbf{R}}-i\mathbf{(RJH-HJR)]Z=\frac{d\ln
\det \mathbf{R}}{dt}.}  \label{46}
\end{equation}%
For any $\mathbf{R}$\textbf{,} $\mathbf{H}$ and $\mathbf{Z}$, Eq. (46)
always holds, then we get Eq. (5) given in the introduction and $\frac{d\ln
\det \mathbf{R}}{dt}=0$. In this way, we derive the Gaussian dynamics
equation in the normal product form. At the same time, there is reason to
believe that $\ln \det \mathbf{R}$ is a constant that does not change with
time. According to the fomula $\det e^{A}=e^{Tr(A)}$, we can obtain $\ln
\det \mathbf{R}=Tr(\ln \mathbf{R})$.

Actually, in Eq. (43), as long as we know that $\frac{i}{2}\sqrt{\det
\mathbf{R}}\frac{\mathbf{E+J}}{2}\frac{\partial }{\partial \mathbf{Z}}[%
\mathbf{HJR\overset{\symbol{126}}{\mathbf{\rho }}\mathbf{(\mathbf{Z})}Z]}$
and $-\frac{i}{2}\sqrt{\det \mathbf{R}}[\mathbf{Z}^{T}\overset{\symbol{126}}{%
\mathbf{\rho }}\mathbf{(\mathbf{Z})RJH]}\overleftarrow{\frac{\partial }{%
\partial \mathbf{Z}^{T}}}\frac{\mathbf{E-J}}{2})$ are all numbers, then,
because of the existence of $\mathbf{E+J}$ and $\mathbf{E-J}$, we can
conclude that $\frac{i}{2}\sqrt{\det \mathbf{R}}\frac{\mathbf{E+J}}{2}\frac{%
\partial }{\partial \mathbf{Z}}[\mathbf{HJR\overset{\symbol{126}}{\mathbf{%
\rho }}\mathbf{(\mathbf{Z})}Z]}$ and $-\frac{i}{2}\sqrt{\det \mathbf{R}}[%
\mathbf{Z}^{T}\overset{\symbol{126}}{\mathbf{\rho }}\mathbf{(\mathbf{Z})RJH]}%
\overleftarrow{\frac{\partial }{\partial \mathbf{Z}^{T}}}\frac{\mathbf{E-J}}{%
2})$ are both equal to $0$. In addition, since $\ln \det \mathbf{R}=Tr(\ln
\mathbf{R})$, then

\begin{eqnarray}
\mathbf{\frac{d\ln \det \mathbf{R}}{dt}} &=&\frac{dTr(\ln \mathbf{R})}{dt}
\label{47} \\
&=&Tr(\overset{\cdot }{\mathbf{R}}\mathbf{R}^{-1})  \notag \\
&=&Tr[i\mathbf{(RJH-HJR)R}^{-1}]  \notag \\
&=&iTr(\mathbf{RJH\mathbf{R}^{-1}-HJ})  \notag \\
&=&i[Tr(\mathbf{RJH\mathbf{R}^{-1})-}Tr\mathbf{(HJ})]  \notag \\
&=&i[Tr(\mathbf{JH)-}Tr\mathbf{(HJ})]  \notag \\
&=&0.  \notag
\end{eqnarray}%
So, we show that if $\overset{\cdot }{\mathbf{R}}=i\mathbf{(RJH-HJR)}$, then
$\mathbf{\frac{d\ln \det \mathbf{R}}{dt}}=0$ naturally satisfies.

Compared with Eq. (2) and Eq. (5), it is not difficult to draw%
\begin{equation}
\mathbf{R(t)=U(t)R(0)U}^{T}\mathbf{(t),}  \label{48}
\end{equation}%
where $\mathbf{U(t)\equiv \exp }(-i\mathbf{JH}t)$. In this way, we get the
solution of Eq. (5) smoothly.

\section{Conclusion}

The time evolution mechanism of Gaussian states is a long-standing and
ever-new topic. This paper mainly provides another mechanism for dealing
with the dynamics of Gaussian states. Different from the previous covariance
mechanism, our work gives the equation for the time evolution of the kernel
matrix $\mathbf{R}$ of Gaussian states in the normal product form, which
provides a new perspective for Gaussian quantum information processing.

The advantage of writing the density matrix of the Gaussian state in the
normal product form is that the specific functional form of the density
matrix under the coherent state representation can be directly given, which
can be done simply by replacing Bosonic operators in the density matrix with
the complex parameters of the coherent state. This processing method will
bring us convenience to solve some problems. For example, for the operator
matrix trace problem, the product of matrices, such as $Tr(\mathbf{AB)}$, is
often encountered. For such problems, we can solve them analytically by
writing $\mathbf{A}$ and $\mathbf{B}$ in the normal product form ($\colon
\overset{\symbol{126}}{\mathbf{A}}\colon $ and $\colon \overset{\symbol{126}}%
{\mathbf{B}}\colon $) and then inserting the completeness of the coherent
state representation ($Tr(\mathbf{AB)=}\diint (d\mathbf{Z}d\mathbf{Z}%
^{\prime })\langle \mathbf{Z}|\colon \overset{\symbol{126}}{\mathbf{A}}%
\colon |\mathbf{Z}^{\prime }\rangle \langle \mathbf{Z}^{\prime }|\colon
\overset{\symbol{126}}{\mathbf{B}}\colon |\mathbf{Z}\rangle $). It is
difficult to solve such problems in a conventional way, especially in the
multi-mode case, and may also have to use numerical methods, while our
method can be solved analytically in principle. Moreover, in the normal
product, we regard Bosonic operators as numbers, so we can perform
integration and differentiation operations without any obstacles, which
cannot be replaced by conventional methods. This processing method
undoubtedly has great potential and has the value of further research and
promotion.

Following the theoretical ideas proposed in this paper, in principle, the
incoherent evolution of the Gaussian state that does not interact with the
environment can be extended to the case in which the system is coherent with
the environment, that is, the Lindblad equation can be solved smoothly,
which will be our follow-up work.

\bigskip \textbf{ACKNOWLEGEMENT: }The work is supported by the School-level
teaching and research project of West Anhui University (Grant wxxy2020047)
and Provincial Teaching and Research Projects of Higher Education
Institutions in Anhui Province (Grant 2021jyxm1666).

\end{document}